\documentclass[final,1p,times]{elsarticle}
\usepackage{amsmath}
\usepackage{graphicx}
\usepackage{amssymb}
\usepackage{amsthm}
\usepackage{lineno}
\usepackage{epsfig}

\journal{Nuclear Physics A} 

\begin{document}

\begin{frontmatter} 

\title{Derivation of transient relativistic fluid dynamics from the Boltzmann equation for a multi-component system}

\author{G.~S.~Denicol${}^{a,b}$ and H.~Niemi${}^{c}$ }

\address{$^{a}$Department of Physics, McGill University, 3600 University Street, Montreal, Quebec, H3A\,2T8, Canada}
\address{$^{b}$Institut f\"ur Theoretische Physik, Johann Wolfgang
Goethe-Universit\"at, Max-von-Laue-Str.\ 1, D-60438 Frankfurt am Main, Germany}
\address{$^{c}$Department of Physics, P.O.Box 35, FI-40014 University of Jyv\"askyl\"a, Finland}

\begin{abstract} 
    We derive the non-equilibrium single-particle momentum distribution function of a hadron resonance gas. We then study the effects that this newly derived expression can have in the freeze-out description of fluid-dynamical models of heavy ion collisions and compare it with the method traditionally employed, the 14-moment approximation.
\end{abstract} 

\end{frontmatter} % do not change

%% linenumbers are useful for reviewing process
%\linenumbers

\section{Introduction}

Fluid-dynamical models have been able to successfully describe the
transverse momentum spectra and azimuthal transverse momentum anisotropies
of particles observed in ultrarelativistic heavy ion collisions. One of the
main ingredients of fluid-dynamical models is the so-called freeze-out
procedure in which the fluid degrees of freedom are matched to kinetic or
particle degrees of freedom. Since experiments at RHIC and LHC measure
particles and not fluid elements, the freeze-out is an essential step
towards the comparison with experimental data.

The Cooper-Frye formalism, usually applied to describe this matching
process, demands the knowledge of microscopic information: the
(single-particle) momentum distribution function of hadrons on the
hypersurface in which freeze-out is performed (usually a constant temperature hypersurface). While this is well known for ideal fluids (Bose-Einstein or
Fermi-Dirac distributions in the local rest frame of the fluid), it has
remained an open problem for viscous fluids. We remark that, so far, Israel
and Stewart's (IS) simple ansatz for \textit{single} component systems is
still being employed in most fluid-dynamical calculations. 

In this case, the
non-equilibrium correction to the momentum distribution of the $i$--th
hadronic species , $\delta f_{\mathbf{k}}^{\left( i\right) }$, is assumed to
be \cite{Teaney:2003kp}
\begin{equation}
\delta f_{\mathbf{k}}^{\left( i\right) }=\varepsilon _{\mu \nu }k_{i}^{\mu
}k_{i}^{\nu },
\end{equation}%
where $k_{i}^{\mu }$ is the four-momentum of the corresponding hadron and $%
\varepsilon _{\mu \nu }$ is an expansion coefficient that should be matched
to the fluid-dynamical variables.

The above expression has two distinct approximations: 1) the momentum
dependence was assumed to be quadratic and 2) the coefficient $\varepsilon
_{\mu \nu }$ was assumed to be the same for all hadronic species, being
matched to the shear stress tensor, $\pi ^{\mu \nu }$, as $\varepsilon _{\mu
\nu }=\pi ^{\mu \nu }/\left[2 \left( \varepsilon +P\right) T^{2}\right] $.
Here, $\varepsilon $, $P$, and $T$ are the energy density, thermodynamic
pressure, and temperature of the fluid. In principle, both these assumptions
are incorrect. The limitations of the first assumption were already
investigated in some works \cite{Schenke}. On the other hand, the second assumption and its domain of validity were just investigated in Ref.~\cite{Molnar:2011kx}. The purpose
of this work is to further elaborate the studies regarding how $\delta f_{\mathbf{k}}^{\left( i\right) }$ actually depends on the different particle species.

\section{Method of moments}

We use the method of moments, as developed in Ref.~\cite{Denicol:2012cn}, to compute the $%
\delta f_{\mathbf{k}}^{\left( i\right) }$ of a multi-component system
without making any \textit{a priori} assumption regarding its momentum
dependence and the particle dependence of the expansion coefficients. First
we factorize $\delta f_{\mathbf{k}}^{\left( i\right) }$ in the following way 
$\delta f_{\mathbf{k}}^{(i)}=f_{0\mathbf{k}}^{(i)}\tilde{f}_{0\mathbf{k}%
}^{(i)}\phi _{\mathbf{k}}^{(i)}$, where $f_{0\mathbf{k}}^{(i)}$ is the local
equilibrium distribution function, $\tilde{f}_{0\mathbf{k}}^{(i)}=1+a\tilde{f%
}_{0\mathbf{k}}^{(i)}$ ($a=-1$/$1$ for fermions/bosons), and $\phi _{\mathbf{%
k}}^{(i)}$ is an out-of-equilibrium contribution. Next, we expand $\phi _{%
\mathbf{k}}^{(i)}$ in terms of its moments using a complete and orthogonal
basis constructed from particle four-momentum, $k_{i}^{\mu }$, and fluid
four-velocity, $u^{\mu }$. As done in Ref.~\cite{Denicol:2012cn}, we use an expansion basis
with two basic ingredients: 1) the irreducible tensors $1,k_{i}^{\left%
\langle \mu \right\rangle },k_{i}^{\left\langle \mu \right. }k_{i}^{\left.
\nu \right\rangle },k_{i}^{\left\langle \mu \right. }k_{i}^{\nu
}k_{i}^{\left. \lambda \right\rangle },\cdots \ $, analogous to the
well-known set of spherical harmonics and constructed by the symmetrized
traceless projection of $k_{i}^{\mu _{1}}\cdots k_{i}^{\mu _{m}}$, i.e., $%
k_{i}^{\left\langle \mu _{1}\right. }\cdots k_{i}^{\left. \mu
_{m}\right\rangle }\equiv \Delta _{\nu _{1}\cdots \nu _{m}}^{\mu _{1}\cdots
\mu _{m}}k_{i}^{\nu _{1}}\cdots k_{i}^{\nu _{m}}$, and 2) the orthonormal
polynomials $P_{i\mathbf{k}}^{\left( n\ell \right)
}=\sum_{r=0}^{n}a_{nr}^{(\ell )i}\left( u_{\mu }k_{i}^{\mu }\right) ^{r}$,
which are equivalent to the associated Laguerre polynomials in the limit of massless,
classical particles \cite{Denicol:2012cn}.

Then, the distribution function becomes,

\begin{equation}
f_{\mathbf{k}}^{(i)}=f_{0\mathbf{k}}^{(i)}+f_{0\mathbf{k}}^{(i)}\tilde{f}_{0%
\mathbf{k}}^{(i)}\sum_{\ell =0}^{\infty }\sum_{n=0}^{\infty }\mathcal{H}_{i%
\mathbf{k}}^{\left( n\ell \right) }\rho _{i,n}^{\mu _{1}\cdots \mu _{\ell
}}k_{i,\mu _{1}}\cdots k_{i,\mu _{\ell }},  \label{expansion}
\end{equation}%
where we introduced the energy-dependent coefficients, $\mathcal{H}_{i%
\mathbf{p}}^{\left( n\ell \right) }\equiv \left[ N_{i}^{\left( \ell \right)
}/\ell !\right] \sum_{m=n}^{\infty }a_{mn}^{(\ell )i}P_{i\mathbf{k}}^{\left(
m\ell \right) }\left( u_{\mu }k_{i}^{\mu }\right) $. The fields $\rho _{i,n}^{\mu _{1}\cdots \mu _{\ell }}$ can be
determined exactly using the orthogonality relations satisfied by the expansion
basis and can be shown to correspond to the irreducible moments of $\delta
f_{\mathbf{k}}^{(i)}$,%
\begin{equation}
\rho _{i,r}^{\mu _{1}\ldots \mu _{\ell }}\equiv \left\langle E_{i\mathbf{k}%
}^{r}k_{i}^{\left\langle \mu _{1}\right. }\ldots k_{i}^{\left. \mu _{\ell
}\right\rangle }\right\rangle _{\delta }\text{, \ \ \ }\left\langle \ldots
\right\rangle _{\delta }=\int dK_{i}\text{ }\left( \ldots \right) \delta f_{%
\mathbf{k}}^{(i)},  \label{Hk}
\end{equation}%
where $g_{i}$ is the degeneracy factor of the $i$--th hadron species and $%
dK_{i}=g_{i}d^{3}\mathbf{k/}\left[ \left( 2\pi \right) ^{3}k_{i}^{0}\right] $%
. As long as this basis is complete, the above expansion fully describes $f_{%
\mathbf{k}}^{(i)}$, no matter how far from equilibrium the system is.

Here, we are interested only on the effects arising from the shear-stress
tensor. For this case, it is enough to fix $\ell =2$ (take only irreducible second-rank
 tensors) in the expansion above, i.e., neglect all scalar terms,
e.g. bulk viscous pressure, irreducible first-rank tensors, e.g. heat flow,
and tensors with rank higher than two (that never appear in fluid dynamics).
The next approximation is the truncation of the expansion in momentum space,
keeping only the term with $n=0$ (with $\ell =2$ already fixed). Then, we
obtain (for classical particles)%
\begin{equation}
f_{\mathbf{k}}^{(i)}=f_{0\mathbf{k}}^{(i)}+\frac{f_{0\mathbf{k}}^{(i)}}{%
2\left( \varepsilon _{i}+P_{i}\right) T^{2}}\pi _{i}^{\mu \nu }k_{i,\mu
}k_{i,\nu }\;.
\end{equation}%
Above, $\pi _{i}^{\mu \nu }=\rho _{i,0}^{\mu \nu }$, $\varepsilon _{i}$, and 
$P_{i}$ are the shear-stress tensor, the energy density, and the
thermodynamic pressure \textit{of the }$i$\textit{--th particle species},
respectively. Note that by keeping only the term with $n=0$ (for $\ell =2$)
we have the same momentum dependence obtained in the IS formula. However,
our expansion coefficients are not independent of the particle type. In this
formalism, this did not have to be assumed, but appeared naturally as a
consequence of the orthogonality relations satisfied by the basis.

In order to apply this expression to describe freeze-out, further
approximations are required. This happens because in fluid dynamics we only
evolve the total shear-stress tensor of the system ($\pi ^{\mu \nu
}=\sum_{i}\pi _{i}^{\mu \nu }$) and do not know, from fluid dynamics itself,
how it divides into the individual shear-stress tensors of each hadron
species ($\pi _{i}^{\mu \nu }$). In order to extract this information, it is
mandatory to know how transient relativistic fluid dynamics emerges from the
underlying microscopic theory.

Basically, transient relativistic fluid dynamics is derived as the long, but
not asymptotically long, time limit of the Boltzmann equation. In this
limit, it is possible to relate the shear-stress tensor of individual
particle species with the total shear stress tensor, $\pi ^{\mu \nu }$, and
the shear tensor, $\sigma ^{\mu \nu }=\partial ^{\left\langle \mu \right.
}u^{\left. \nu \right\rangle }$, of the system. This was done in Ref.~\cite{Denicol:2012cn}, for single component systems, and can be easily
extended to multi-component systems. Here, we just write down the solution
and leave the detailed derivation of this relation to a future work. The solution is, 
\begin{equation}
\pi _{i}^{\mu \nu }=\frac{\Omega ^{i0}}{\sum_{j}\Omega ^{j0}}\pi ^{\mu \nu
}+2\left[ \eta _{i}-\eta \frac{\Omega ^{i0}}{\sum_{j}\Omega ^{j0}}\right]
\sigma ^{\mu \nu }\equiv \alpha _{i}\pi ^{\mu \nu }+\beta _{i}\sigma ^{\mu
\nu },  \label{EquationN}
\end{equation}%
where $\eta _{i}$ is the shear viscosity of the $i$--th species, while $\eta
=\sum_{i}\eta _{i}$ is the total shear viscosity. The matrices $\Omega $ are
defined in such a way as to diagonalize the collision integral $\mathcal{M}$%
, $\Omega ^{-1}\mathcal{M}\Omega =\mathrm{diag}(\chi _{1},\cdots ,\chi _{N})$%
, with $\chi _{i}$ being the eigenvalues of $\mathcal{M}$. With the
truncation in momentum assumed above and considering only elastic 2-to-2
collisions between \textit{classical} particles, $\mathcal{M}$ has the
following simple form 
\begin{gather}
\mathcal{M}^{ij}=\mathcal{A}^{i}\delta ^{ij}+\mathcal{C}^{ij}, \\
\mathcal{A}^{i}=\frac{1}{5}\sum_{j=1}^{N_{\mathrm{hadr.}}}\int
dK_{i}dK_{j}^{\prime }dP_{i}dP_{j}^{\prime }\gamma _{ij}W_{\mathbf{pp}%
^{\prime }-\mathbf{kk}^{\prime }}^{ij}f_{i\mathbf{k}}^{\left( 0\right) }f_{j%
\mathbf{k}^{\prime }}^{\left( 0\right) }\left( u_{\lambda }k_{i}^{\lambda
}\right) ^{-1}k_{i}^{\left\langle \mu \right. }k_{i}^{\left. \nu
\right\rangle }\left[ p_{i\left\langle \mu \right. }p_{i\left. \nu
\right\rangle }-k_{i\left\langle \mu \right. }k_{i\left. \nu \right\rangle }%
\right] , \\
\mathcal{C}^{ij}=\frac{1}{5}\int dK_{i}dK_{j}^{\prime }dP_{i}dP_{j}^{\prime
}\gamma _{ij}W_{\mathbf{pp}^{\prime }-\mathbf{kk}^{\prime }}^{ij}f_{i\mathbf{%
k}}^{\left( 0\right) }f_{j\mathbf{k}^{\prime }}^{\left( 0\right) }\left(
u_{\lambda }k_{i}^{\lambda }\right) ^{-1}k_{i}^{\left\langle \mu \right.
}k_{i}^{\left. \nu \right\rangle }\left[ p_{j\left\langle \mu \right.
}^{\prime }p_{j\left. \nu \right\rangle }^{\prime }-k_{j\left\langle \mu
\right. }^{\prime }k_{j\left. \nu \right\rangle }^{\prime }\right] ,
\end{gather}%
where $\gamma ^{ij}=1-\left( 1/2\right) \delta ^{ij}$ and $W_{\mathbf{pp}%
^{\prime }-\mathbf{kk}^{\prime }}^{ij}$ is the transition rate.

\section{Simple model of hadrons}

In order to compute the coefficients $\alpha _{i}$ and $\beta _{i}$
appearing in Eq.~\ref{EquationN} we must provide
a set of hadronic cross sections. In this work, we estimate these coefficients for the first time
using a simple hadronic model, in which all hadrons have the same constant
cross section, $\sigma _{ij}=$ $30$ mb. We consider only elastic collisions
between the hadrons and include all hadrons up to a mass of $1.8$ GeV. In
this case, $\alpha _{i}$ and $\beta _{i}$ are actually independent of the
value chosen for the cross section and all the difference between the
various hadrons species originate solely from their different masses. In
this simple scheme, we computed $\alpha _{i}$ and $\beta _{i}$ for all
hadron species up to $1.8$ GeV and included all the decays of unstable resonances.

In order to probe the effect this more sophisticated $\delta f_{\mathbf{k}%
}^{(i)}$ can have on heavy-ion observables, we computed the elliptic flow of
pions, kaons, and protons using the fluid-dynamical model applied in Ref.~\cite{Niemi:2012ry}, with exactly the same parameters (GLmix initialization, HH-HQ
shear viscosity parametrization), but considering several choices of $\delta
f_{\mathbf{k}}^{(i)}$. The resutls are shown in Figs.1. The different lines
correspond to the following cases: 1) solid line uses Israel-Stewart's
ansatz, 2) dotted line uses the $\delta f_{\mathbf{k}}^{(i)}$ derived in
this paper, and 3) dashed line uses the $\delta f_{\mathbf{k}}^{(i)}$
derived in this paper in the Navier-Stokes limit ($\pi ^{\mu \nu }=2\eta
\sigma ^{\mu \nu }$).

\begin{figure}[!h]
% \vspace{-0.5cm}
\hspace{-0.0cm} \epsfysize 3.4cm \epsfbox{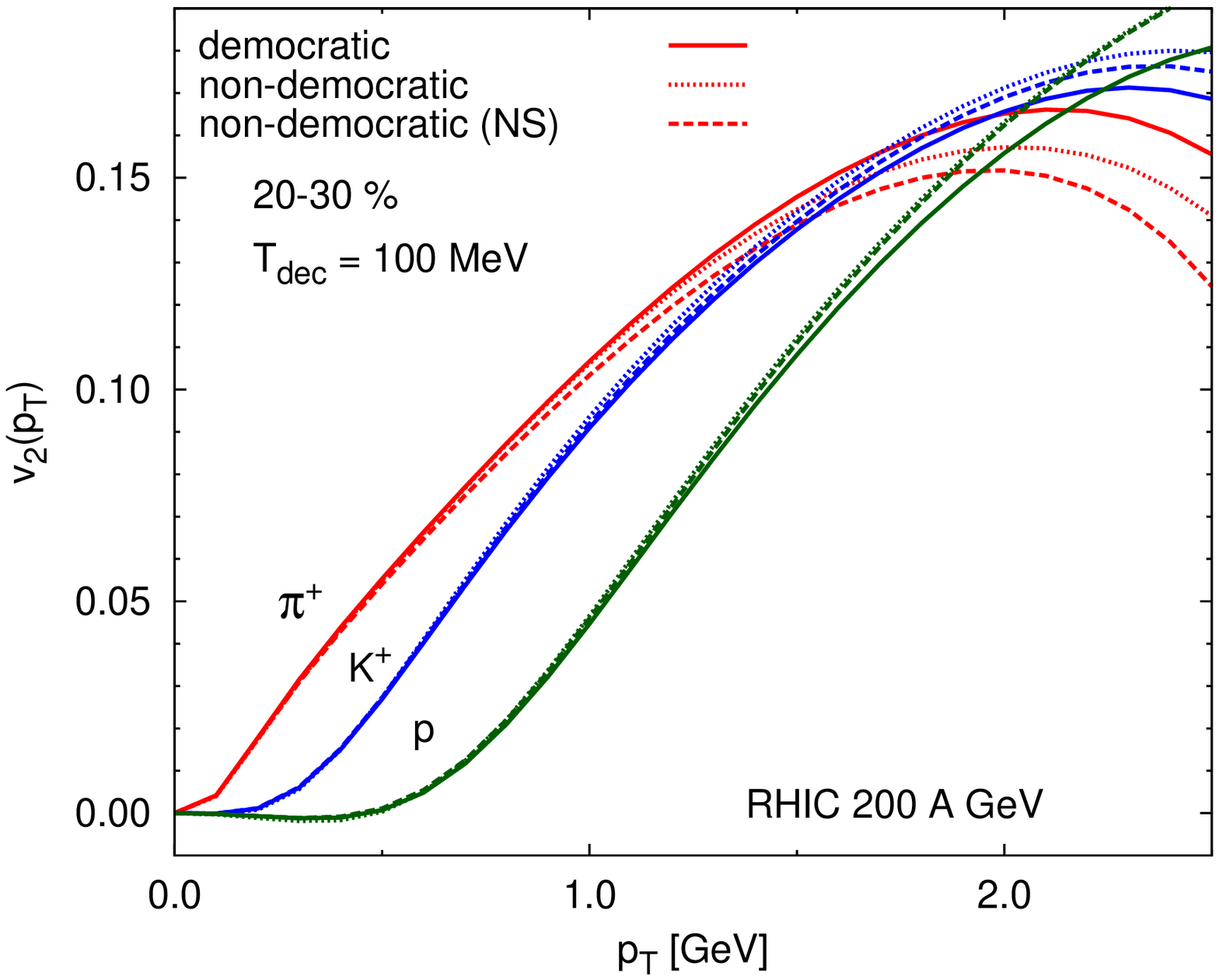} \hspace{%
0.0cm} \hspace{-0.0cm}\epsfysize 3.4cm %
\epsfbox{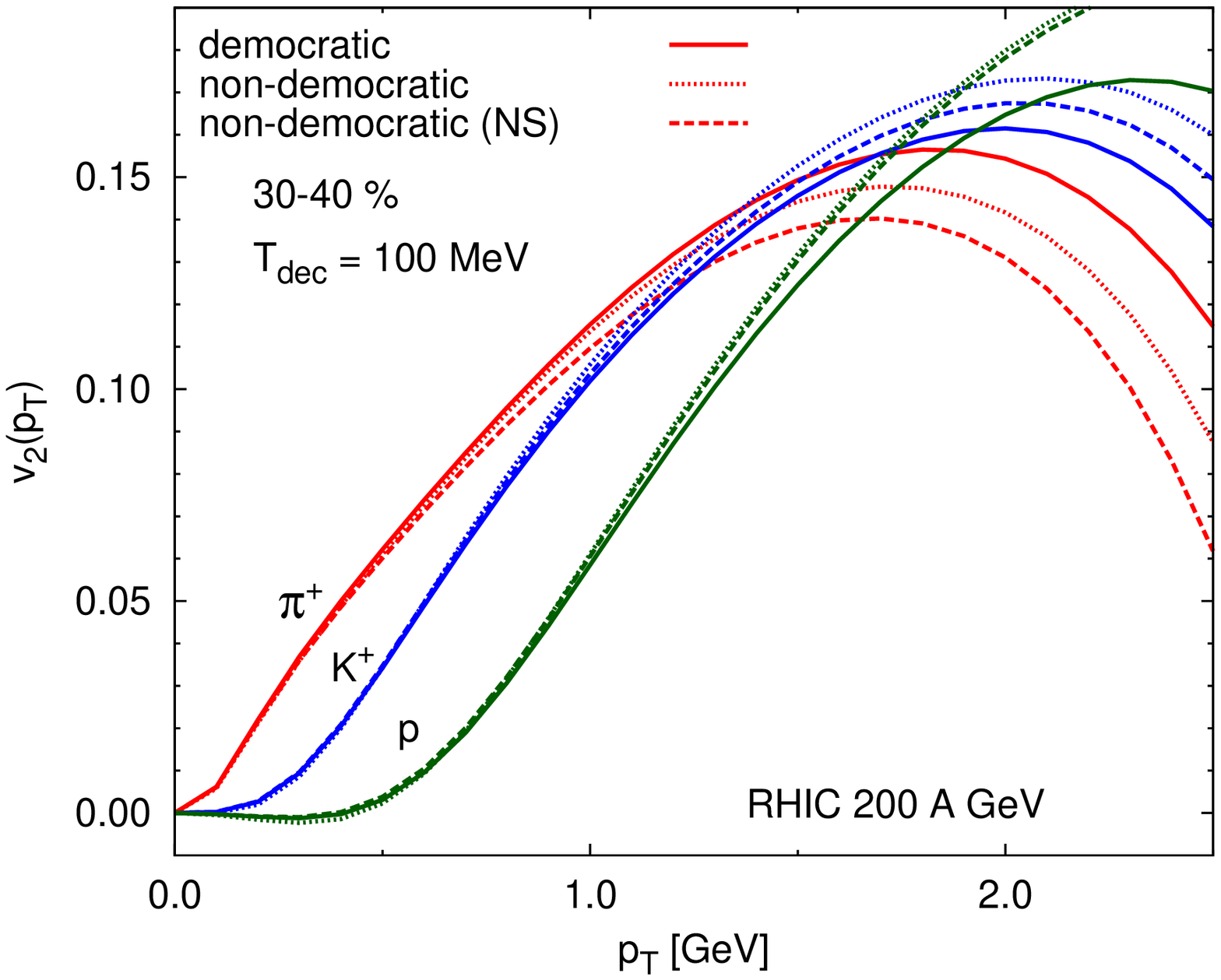} \hspace{ 0.0cm}\epsfysize 3.4cm %
\epsfbox{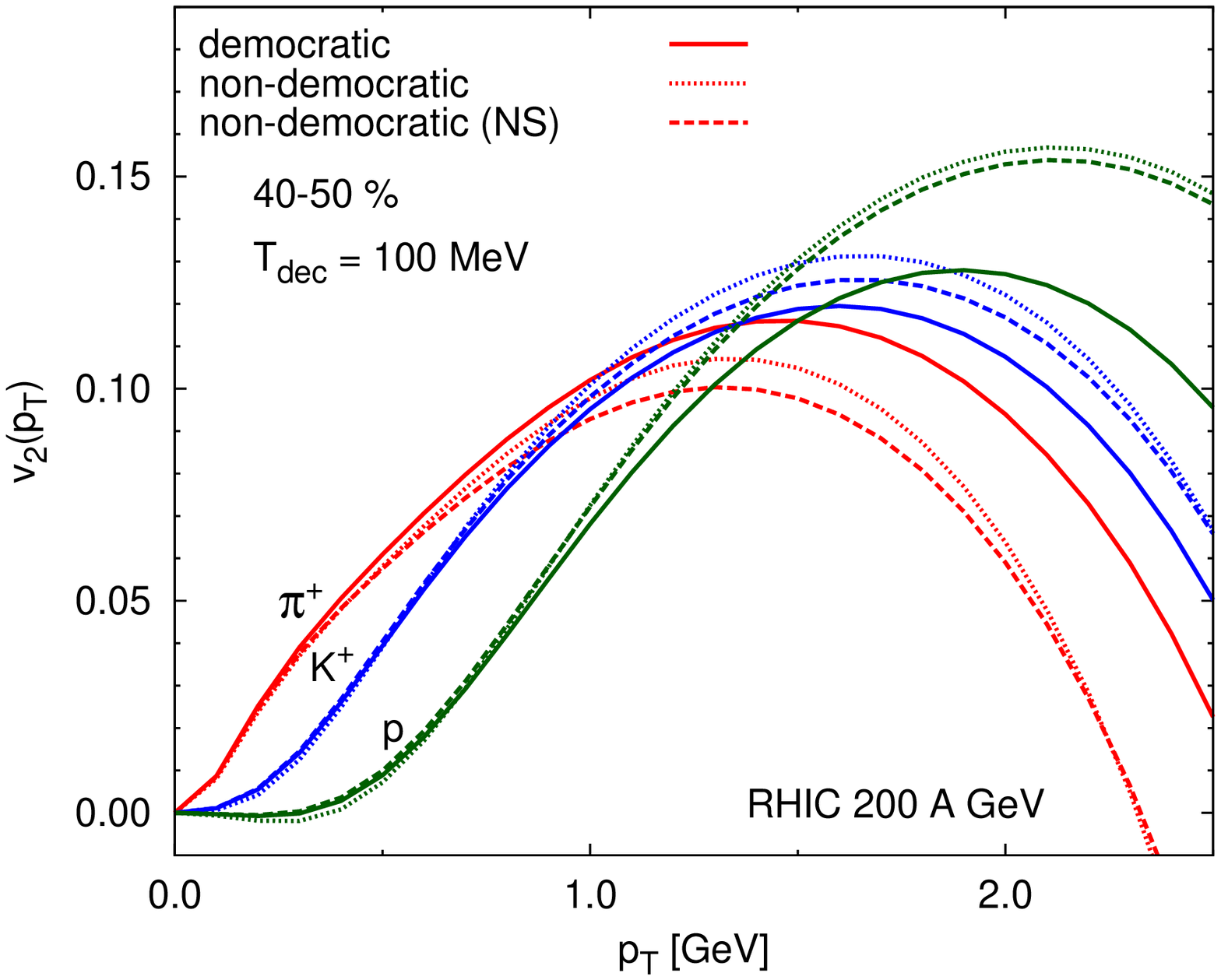} % \vspace{-0.5cm}
\caption{Differential elliptic flow for pions, kaons, and protons for three
different centrality classes, (a) $20-30$ \%, (b) $30-40$ \%, (c) $40-50$
\%, and various choices of $\protect\delta f_{\mathbf{k}}^{(i)}$.}
\label{fig:e2v2_20_30}
\end{figure}

We see that the particle dependence of the coefficients do not have a strong
effect on the differential elliptic flow, which looks very similar to the
one computed using Israel-Stewart's ansatz. At very non-central collisions,
the difference is larger, but even then, cannot be considered as a crucial
effect. Note that one of the reasons that makes this difference small is
that we are not in the Navier-Stokes limit at freeze-out: the elliptic flow
computed assuming the Navier-Stokes limit actually deviates more strongly from the
one computed using Israel-Stewart's ansatz. Furthermore, note that there is
a qualitative particle dependence in our result, the elliptic flow
of pions is below the one computed with Israel-Stewart's ansatz
while the elliptic flow of kaons and protons are always above.

We remark that these results arise from an over simplified hadronic
description and this insensitivity to the particle species might be coming just
from the simple choice of hadronic cross sections assumed. In future
calculations, we will come back to this discussion using more realistic
cross sections for the hadrons and also including inelastic collisions between the hadrons as
well.

%\section*{References}

The work of H.N.\ was supported by the Academy of Finland, Project No.~133005. G.S.D thanks D.~Molnar for helpful discussions.

\end{document}